\documentclass[10pt,letterpaper]{article}

\usepackage[top=0.85in,left=2.75in,footskip=0.75in]{geometry}
\usepackage{subcaption}
\usepackage{changepage}

\usepackage[utf8]{inputenc}

\usepackage{textcomp,marvosym}

\usepackage{fixltx2e}

\usepackage{amsmath,amssymb}

\usepackage{float}

\usepackage{filecontents}

\usepackage{cite}

\usepackage[right]{lineno}

\usepackage{microtype}
\DisableLigatures[f]{encoding = *, family = * }

\usepackage{rotating}

\raggedright
\usepackage[aboveskip=1pt,labelfont=bf,labelsep=period,justification=raggedright,singlelinecheck=off]{caption}

\makeatletter
\renewcommand{\@biblabel}[1]{\quad#1.}
\makeatother

\date{}

\usepackage{lastpage,fancyhdr,graphicx}
\usepackage{epstopdf}
\usepackage[final]{changes}

\begin{document}
\vspace*{0.35in}

\begin{flushleft}
{\Large
\textbf\newline{Structure of 311 Service Requests as a Signature of Urban Location
}
\newline
\renewcommand{\thefootnote}{\fnsymbol{footnote}}
\\
Lingjing Wang\textsuperscript{1,2},
Cheng Qian\textsuperscript{1,2},
Constantine Kontokosta\textsuperscript{1,2},
Stanislav Sobolevsky\textsuperscript{1,*}

\bigskip
\textsuperscript{1} \added{Center for Urban Science and Progress, New York University, Brooklyn, New York, United States of America}\\
\textsuperscript{2} \added{Tandon School of Engineering, New York University, Brooklyn, New York, United States of America}\\
\textsuperscript{*} \added{Correspondence should be addressed: sobolevsky@nyu.edu}
\\
\bigskip






}

\end{flushleft}

\section*{Abstract}

While urban systems demonstrate high spatial heterogeneity, many urban planning, economic and political decisions heavily rely on a deep understanding of local neighborhood contexts. We show that the structure of 311 Service Requests enables one possible way of building a unique signature of the local urban context, thus being able to serve as a low-cost decision support tool for urban stakeholders. Considering examples of New York City, Boston and Chicago, we demonstrate how 311 Service Requests recorded and categorized by type in each neighborhood can be utilized to generate a meaningful classification of locations across the city, based on distinctive socioeconomic profiles. Moreover, the 311-based classification of urban neighborhoods can present sufficient information to model various socioeconomic features. Finally, we show that these characteristics are capable of predicting future trends in comparative local real estate prices. We demonstrate 311 Service Requests data can be used to monitor and predict socioeconomic performance of urban neighborhoods, allowing urban stakeholders to quantify the impacts of their interventions.



\section*{Introduction}

Cities can be seen as a complex system composed of multiple layers of activity and interactions across various urban domains; therefore, discovering a parsimonious description of urban function is quite difficult\cite{batty2008size,bettencourt2010urbscaling, bettencourt2013origins, arcaute2013citybound}. However, urban planners, policy makers and other types of urban stakeholders, including businesses and investors, could benefit from an intuitive proxy of neighborhood conditions across the city and over time\cite{const2015, maimon_datamining, townsend2013}. At the same time, such simple indicators could provide not only valuable information to support urban decision-making, but also to accelerate the scalability of successful approaches and practices across different neighborhood and cities, as urban scaling patterns have become an increasing topic of interest \cite{powell2007food, bettencourt2007growth, albeverio2008, sobolevsky2014mining, sobolevsky2015scaling}. As the volume and heterogeneity of urban data have increased, machine learning has become a viable tool for enhancing our knowledge of urban space and in developing predictive analytics to inform city management and policy\cite{nelder,bb2003, macqueen1967kmeans, rousseeuw1987}.

The non-trivial challenge is to identify a consistent, quantifiable metric that provides comprehensive insights across multiple layers of urban operations and planning \cite{allwinkle2011} and to locate readily-available data to support its implementation across a range of cities. 
Fortunately, urban data collected by various agencies and companies provide an opportunity to respond to this challenge\cite{batty2012, bettencourt2014bigdata}. In the age of ubiquitous digital media, numerous aspects of human activity are being analyzed by means of their digital footprints, such as mobile call records \cite{ girardin2008digital, gonzalez2008uih, quercia2010rse, sobolevsky2013delineating, amini2014impact, reades2007, const2016}, vehicle GPS traces \cite{santi2014}, bank card transactions \cite{sobolevsky2016prism, shen2014, scholnick2013}, payment patterns\cite{boeschoten1998, bounie2006, hayhoe2000}, smart card usage \cite{bagchi2005, lathia2012, chan1999frauddetection, rysman2007}, or social media activity \cite{szell2013, frank2013happiness, hawelka2014, paldino2015, lenormand_bcards_2014}.
Such data sets have been successfully applied to investigate urban \cite{louail2014citystruct} and regional structure \cite{ratti2010redrawing, sobolevsky2013delineating}, land use \cite{grauwin2014towards, pei2014new}, financial activities\cite{sobolevsky2014money}, mobility \cite{noulas2012tale, kung2014exploring}, or well-being \cite{lathia2012, sobolevsky2015predicting}.

However, one of the major limitations to widespread adoption of such analytics in the practice of urban management and planning is the extreme heterogeneity of the data coverage: different types of data are available for different areas and periods of time, which undermine efforts to develop universal and reliable analytic approaches. Privacy considerations are another significant issue that create additional practical and legal obstacles, restricting data access and preventing their use out of a concern for confidentiality\cite{lane2014, christin2011, belanger2011, krontiris2010}.

Increasingly, cities are introducing systems to collect service requests and complaints from their citizens.  These data, commonly referred to as 311 requests, reflect a wide range of concerns raised by city residents and visitors, offering a unique indicator of local urban function, condition, and service level. In many cities, 311 requests are publicly available through city-managed open data platforms as part of a broader movement in local government to increase transparency and good governance \cite{walker2013}. Although potentially biased by the self-reported nature of the requests and complaints, these data provide a comparable measure of perceived local quality of life across space and time.

In this article, we develop a method for classifying urban locations based on the categorical and temporal structure of 311 Service Requests for a given neighborhood, exploring whether these spatio-temporal patterns can reveal characteristic signatures of the area. For New York, Boston, and Chicago, we present applications of this new urban classifier for predicting socioeconomic and demographic characteristics of a neighborhood and estimating the economic performance and well-being of a defined spatial agglomeration. The paper begins with a discussion of the data and methodology, followed by specific use cases relating to demographics and real estate values, and concluding with opportunities for future research.


\section{Materials and Resources}
\subsection{The 311 data}

{ 311 service request and complaint data are being collected across more than 30 cities in the United States, including New York, Boston and Chicago. Through the 311 system, local government agencies offer non-emergency services to residents, visitors and businesses and respond to reported service disruptions, unsafe conditions, or quality-of-life disturbances. These 311 service requests and complaints cover a wide range of concerns, including, but not limited to, noise, building heat outages, rodent sightings, etc. Thus, these data serves as an extremely useful resource in understanding the delivery of critical city services and neighborhood conditions.

We explore the 311 datasets from New York, Chicago, and Boston as major urban centers where 311 systems are in place and commonly used. We consider a time frame between 2012 and 2015 during which the data are available for all three cities selected. In table 1, we provide descriptive statistics of the data. Note that the number of total requests has been increasing from 2012 to 2015 in each city. Conceivably, the number of requests in New York City (which now approaches 2 million per year) is higher than the others because of its population size. However, Boston has a substantially smaller number of requests compared to the similar-sized city of Chicago, which shows the discrepancies in the use of the system across cities. Unfortunately, each city uses a different complaint/request coding convention, thus there is little consistency in the classification of particular complaint types. This fact raises certain difficulties for analysis between cities, a common challenge in comparative urban analytics given the lack of data standardization. For example, in 2015, New York City's 311 data are categorized into 182 types, where Chicago has only 12. Even within a particular city, request categories are subject to change over time, especially in NYC where only approximately 70\% of the entire service request activity belong to common categories present in all four years. Additional adjustments are needed to re-classify complaint types into standardized categories across the different cities and over the time period of the analysis.

}

The original data set provided by 311 Services contains one record for each customer's call. For most cities, these records include information such as: service request type, service request open/close time and date and location(longitude and latitude). Therefore, for any given time period and area(census tract area/zipcode area), we can aggregate the 311 service requests and group by type.

\begin{table}[ht]
\centering
\begin{tabular}{|*{10}{c|}}  
\hline
\multicolumn{1}{|c}{Year} & \multicolumn{9}{|c|}{New York City}\\ \hline

\multicolumn{1}{|c}{} & \multicolumn{3}{|c|}{Total Requests}& \multicolumn{3}{|c|}{Requests Categories}& \multicolumn{3}{|c|}{Share of common categories' activity}\\ \hline

\multicolumn{1}{|c}{2012} & \multicolumn{3}{|c|}{1414392}& \multicolumn{3}{|c|}{165}& \multicolumn{3}{|c|}{0.69}\\ \hline

\multicolumn{1}{|c}{2013}  & \multicolumn{3}{|c|}{1431729}& \multicolumn{3}{|c|}{162}& \multicolumn{3}{|c|}{0.69}\\ \hline

\multicolumn{1}{|c}{2014} & \multicolumn{3}{|c|}{1654913}& \multicolumn{3}{|c|}{179}& \multicolumn{3}{|c|}{0.73}\\ \hline

\multicolumn{1}{|c}{2015} & \multicolumn{3}{|c|}{1806560}& \multicolumn{3}{|c|}{182}& \multicolumn{3}{|c|}{0.73}\\ \hline

\multicolumn{1}{|c}{Year} & \multicolumn{9}{|c|}{Chicago}\\ \hline
\multicolumn{1}{|c}{} & \multicolumn{3}{|c|}{Total Requests}& \multicolumn{3}{|c|}{Requests Types}& \multicolumn{3}{|c|}{Share of common categories' activity}\\ \hline

\multicolumn{1}{|c}{2012} & \multicolumn{3}{|c|}{478532}& \multicolumn{3}{|c|}{13}& \multicolumn{3}{|c|}{0.85}\\ \hline

\multicolumn{1}{|c}{2013}  & \multicolumn{3}{|c|}{507956}& \multicolumn{3}{|c|}{14}& \multicolumn{3}{|c|}{0.82}\\ \hline

\multicolumn{1}{|c}{2014} & \multicolumn{3}{|c|}{515258}& \multicolumn{3}{|c|}{14}& \multicolumn{3}{|c|}{0.82}\\ \hline

\multicolumn{1}{|c}{2015} & \multicolumn{3}{|c|}{568576}& \multicolumn{3}{|c|}{12}& \multicolumn{3}{|c|}{0.9}\\ \hline

\multicolumn{1}{|c}{Year} & \multicolumn{9}{|c|}{Boston}\\ \hline
\multicolumn{1}{|c}{} & \multicolumn{3}{|c|}{Total Requests}& \multicolumn{3}{|c|}{Requests Types}& \multicolumn{3}{|c|}{Share of common categories' activity}\\ \hline

\multicolumn{1}{|c}{2012} & \multicolumn{3}{|c|}{92855}& \multicolumn{3}{|c|}{155}& \multicolumn{3}{|c|}{1}\\ \hline

\multicolumn{1}{|c}{2013}  & \multicolumn{3}{|c|}{112727}& \multicolumn{3}{|c|}{165}& \multicolumn{3}{|c|}{0.99}\\ \hline

\multicolumn{1}{|c}{2014} & \multicolumn{3}{|c|}{112785}& \multicolumn{3}{|c|}{183}& \multicolumn{3}{|c|}{0.96}\\ \hline

\multicolumn{1}{|c}{2015} & \multicolumn{3}{|c|}{161498}& \multicolumn{3}{|c|}{180}& \multicolumn{3}{|c|}{0.83}\\ \hline

\end{tabular}

\caption{General properties of the 311 data for NYC, Chicago and Boston} 
\end{table} 

\subsection{Demographic and socio-economic  data}

As we are attempting to use 311 data as a proxy for the socioeconomic characteristics and real estate values of urban neighborhoods, ground-truth data are needed to train and validate our models. For socioeconomic and demographic features, we use data from the U.S. Census 2014 American Community Survey (ACS). For real estate values, we collect housing price data from the online real estate listing site Zillow. Both are described below.

\subsubsection{2014 census data}

The 2014 ACS contains survey data on a number of socieconomic and demographic variables, at the spatial aggregation of the Census Block. For this analysis, we have selected common features representing important phenomena in population diversity, education, and income and employment. For example, our selection covers the number of population in the following categories: "Non-Hispanic White", "African-American", "Asian", "High school degree", "College degree", "Graduate degree", "Uninsured ratio", "Unemployment ratio", "Poverty ratio", and mean for "Income (all)", "Income of No Family", "Income of Families" and "Income of Households".

One important consideration is the level of spatial aggregation for this analysis. Having considered zip code, census tract and census block, we decided to proceed with census tracts providing the best trade-off between spatial granularity, in terms of having a sufficient number of sub-areas within each city, and having a statistically significant sample of 311 complaints for each areal unit. In Boston and Chicago, there are too few zipcodes within in each city to create a useful sample, and there is not a significant density of 311 complaints at the census block level (please refer to SI4 for details). In addition, given the survey methodology of the ACS data, census block data include non-trivial margins-of-error for each variable. 



\subsubsection{Zillow Housing price}
One important indicator of local economic conditions is housing prices \cite{const2012}. We utilize Zillow housing price data that contain monthly average residential real estate sales prices by zip code. Although housing prices are a lagging indicator of neighborhood economic strength, since recorded sales occur as much as two to more than six months after a contract is signed, we use these values as one of the targets for our 311 predictions. Our spatial level of analysis will be the zipcode, rather than census tracts, given the coverage area of the Zillow aggregate data.

\subsubsection{Normalization method and some notations}

In order to better compare various areas, the census data need to be normalized. Take income per capita and population with bachelor degree for example. Firstly, these two features have different measurement units (dollars versus number of people). Secondly, this number can be affected by the area's total population. For an area with high population, there should be a higher possibility to have higher population with bachelor degree. Therefore, the normalization process is important in order to compare different features and different areas with heterogeneous population. For our analysis, we normalize our census tract data set in the following way. 

Let $p_i$ be the total population in census tract, while $v_i$ denotes one feature recorded in the same census tract $i$, for example, "the total population who holds graduate degree in census tract $i$". Next, we normalize it by defining

$$y_i=\frac{v_i  p_i}{\sum_{j\in{\Omega}}v_j  p_j}$$  \\\

We define $\Omega$ as a set of all census tracts in New York City.

In section 2, we use 311 complaint frequency categorized by census tracts to cluster and investigate the difference in local socioeconomic features $y$. In section 3 we use machine learning regression models to predict these features $y$ using normalized 311 data .

\section{Classification based on 311 service categories}

In order to get initial insights on the usage of 311 service across the considered cities, we define for each census tract a 311 service request signature - a vector of the relative frequencies of 311 requests of different types. Specifically, let the total number of service requests of each type $t$ within an area $a$ be $s(a,t)$ and let $s(a)=\sum_t s(a,t)$ be the total number of service requests in the area $a$. Then a vector $S(a)=(s(a,t)/s(a), t=1..T)$, where $T$ is the total number of service request types, serves as a signature of the location's aggregated 311 service request behavior. The vector $S$ highlights the primary reasons for service requests or complaints in the specific area, as well as allowing for straightforward comparison across tracts and cities.

Signatures $S(a)$ serve as unique characteristics of each location $a$, and we would expect similar spatio-temporal patterns to emerge in 311 service requests across a city or cities. Our hypothesis here is that these similarities also suggest  similarities in the socioeconomic characteristics of the areas. In order to explore this further, we apply a $k$-means clustering approach to the set of multi-dimensional vectors $S(a)$. In order to ensure we get an optimal clustering we run the $k$-mean 100 times, selecting the best solution in terms of cumulative square sum of distances from centroids.

One crucial step in this approach is to pick up an appropriate number of clusters to consider. For that purpose we have evaluated the clustering model with both Silhouette method and Elbow method. While different methods give a slightly different optimal number of clusters for the cities in our sample, in most cases it is within a range of two to four clusters. Given the socioeconomic diversity across neighborhoods in the selected cities, we determine that a minimum of four clusters is an appropriate value. Readers can find more details in SI.

\begin{figure*}[h]
\centering
\includegraphics[width=13cm, height=13cm]{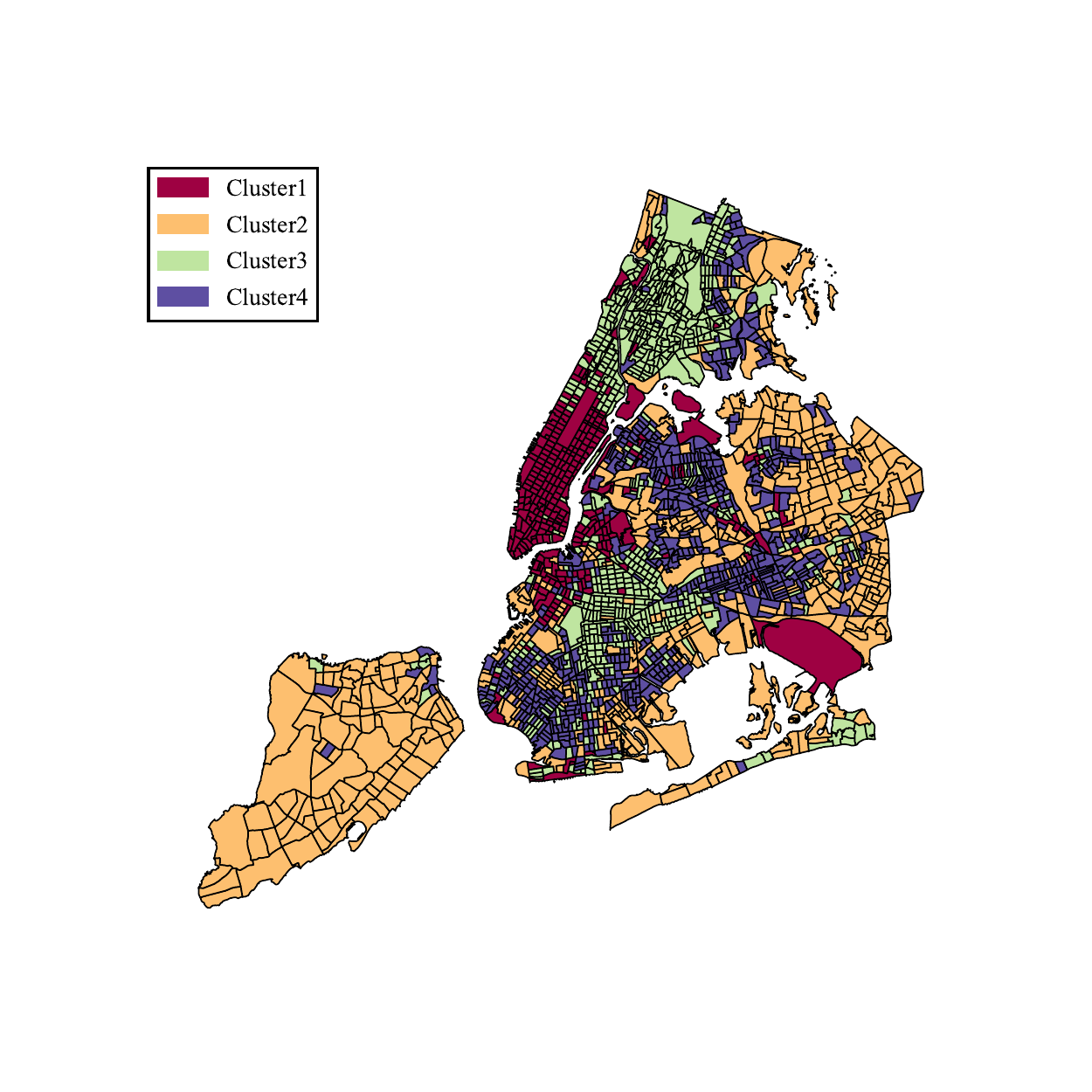}
\caption{\label{fig::category_classification} Classification of urban locations based on the categorical structure of the 311 requests.}
\end{figure*}

We consider NYC first. In Figure 1, we see below with approximate 2000 census tracts divided into four clusters based on our clustering results. Midtown Manhattan, downtown Brooklyn and several outliers such as JFK and LGA airports belong to cluster 1; Staten Island and eastern Brooklyn/Queens constitute cluster 2; Northern Manhattan, the Bronx, and central Brooklyn are included in cluster 3; and Southern Brooklyn, Flushing and some eastern parts of Bronx comprise cluster 4. 

\begin{figure*}[h]
\centering
\includegraphics[width=14cm, height=7cm]{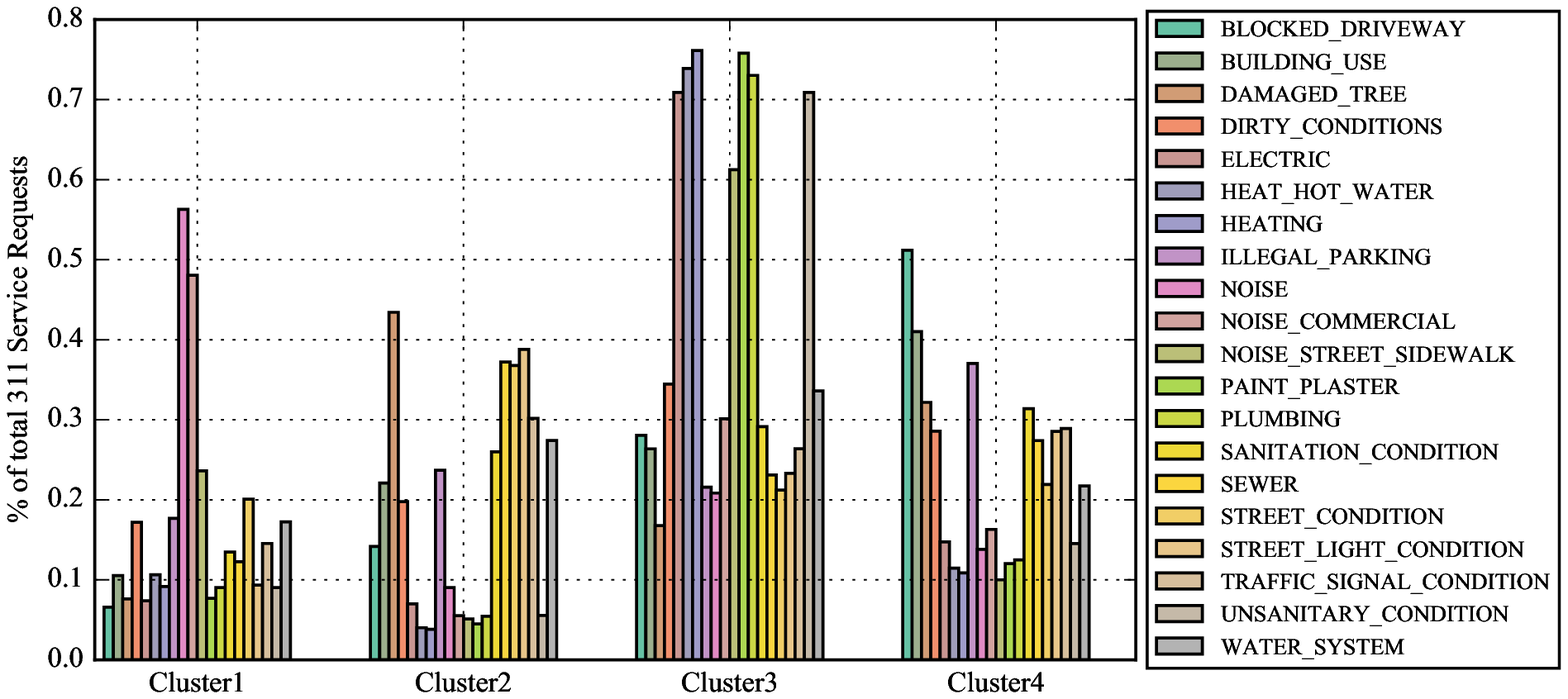}
\caption{\label{fig::category_classification} Top 20 requests distribution among clusters.}
\end{figure*}

In order to evaluate how different each cluster is with respect to the nature of 311 service requests, (see figure 2) we present the distribution of top service requests over the four clusters. We observe clear variation in this distribution. For example, complaints/requests within cluster 1 more often report noise concerns than others, cluster 2 experiences more issues relating to residential heating, cluster 3 has the highest relative complaints about blocked driveways, while cluster 4 reports concerns about street conditions. 

Similarly, we repeat the same clustering process for Chicago and Boston and the clustering results for census tracts in those cities are shown in Figure 3. 


\begin{figure*}[h]
\begin{subfigure}{.5\textwidth}
  \centering
  \includegraphics[width=.8\linewidth]{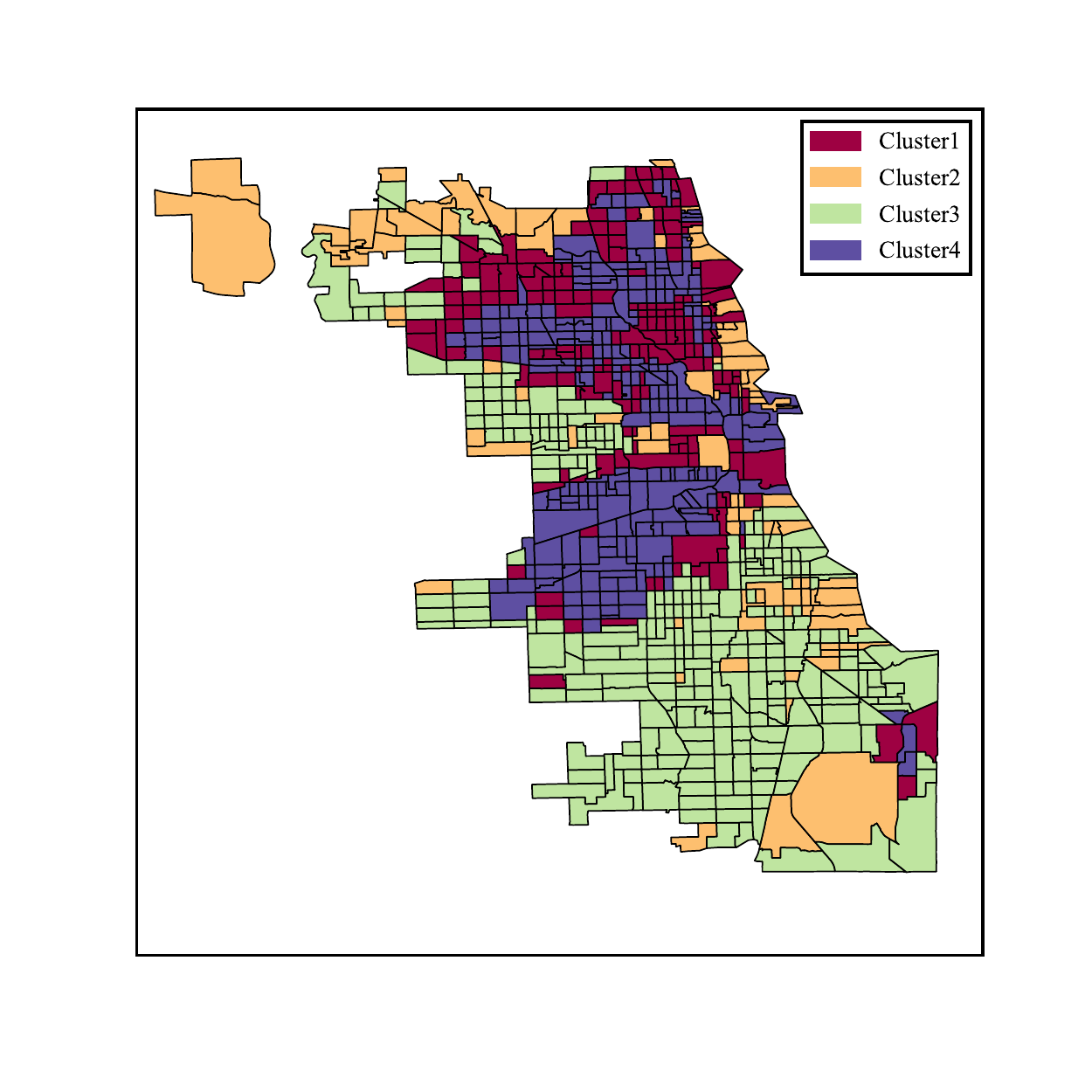}
  \caption{Chicago}
  \label{fig:sfig1}
\end{subfigure}%
\begin{subfigure}{.5\textwidth}
  \centering
  \includegraphics[width=.8\linewidth]{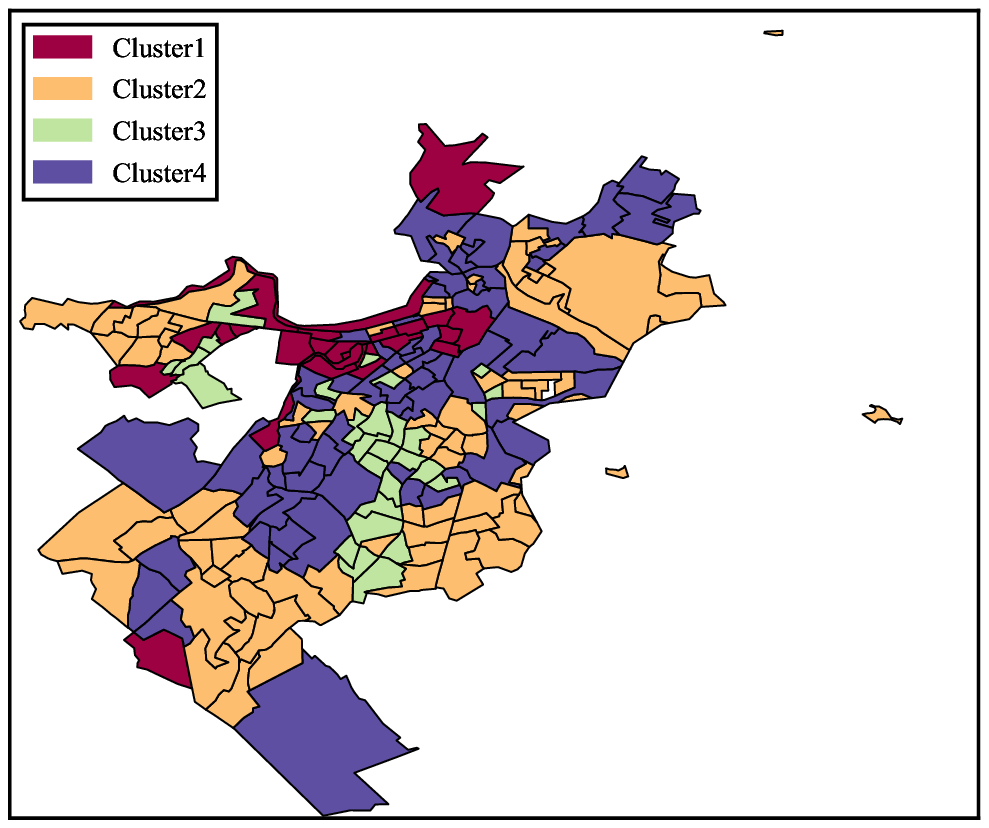}
  \caption{Boston}
  \label{fig:sfig2}
\end{subfigure}

\caption{\label{fig::category_classification} Classification of urban locations based on the categorical structure of the 311 service requests for Chicago and Boston.}
\label{fig:fig}
\end{figure*}

\section{Socioeconomic features among clusters}

Given knowledge of the local spatial contexts for the analyzed cities, the clusters that emerge make certain intuitive sense. However, in order to quantitatively address the hypothesis formulated in the previous section - that similarities in local 311 service request signatures also imply similarities in the socioeconomic profiles of those areas - here we summarize and analyze the socioeconomic characteristics for each of the discovered clusters. 

Recall that thus far the clustering results are obtained based on the 311 service requests frequency alone with no socioeconomic information considered. Next we summarize 14 socioeconomic features and compare the normalized mean level for each feature in each of the considered clusters. The results for our three cities are presented in the radar plots in Figures 4-6. From the output, we can see that the socioeconomic features among the defined clusters are quite distinctive.

Take NYC for example:

\begin{itemize}
    \item  Education and Income: People with higher levels of education (with graduate degree and above) are found in cluster 1, which, as expected, also has highest income level. Cluster 3 appears to show the opposite results.
    \item  Racial diversity: There are above average concentrations of Non-Hispanic Whites living in clusters 1 and 2, of Asian origin in cluster 4, and African-American populations in cluster 3.
\end{itemize}



Similarly we have (for both Chicago and Boston):
\begin{itemize}
    \item  Cluster 1 has the highest income and education level, while cluster 3 is the lowest.
    \item   Cluster 2 is predominantly Asian and African-Americans, while Non-Hispanic Whites tend to live in clusters 1 and 4.
\end{itemize}


The observations above provide some evidence for our hypothesis, revealing links between socioeconomic features and 311 service request data structure. Indeed, while the clustering is performed based on the 311 data alone, the socioeconomic features happen to be quite distinctive among the clusters. Of course this only reveals the existence of a certain relation in principle, which might not be that practical. However this gives rise to another hypothesis - can one use 311 service request data to actually model socioeconomic features at the local scale?

\begin{figure}[H]
\begin{subfigure}{.5\textwidth}
  \centering
  \includegraphics[height=7cm]{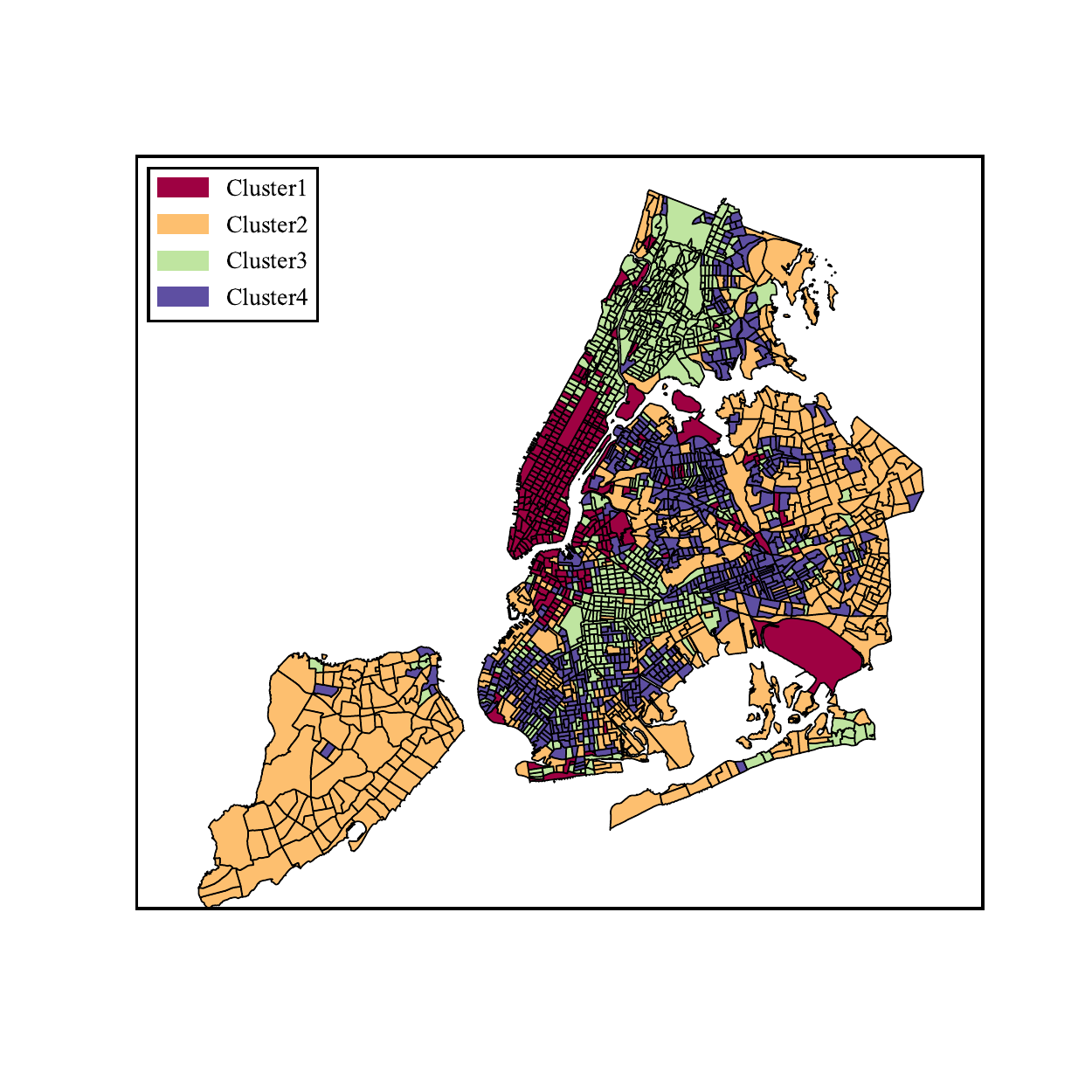}
  \label{fig:sfig2}
\end{subfigure}%
\begin{subfigure}{0.5\textwidth}
  \centering
  \includegraphics[height=6cm]{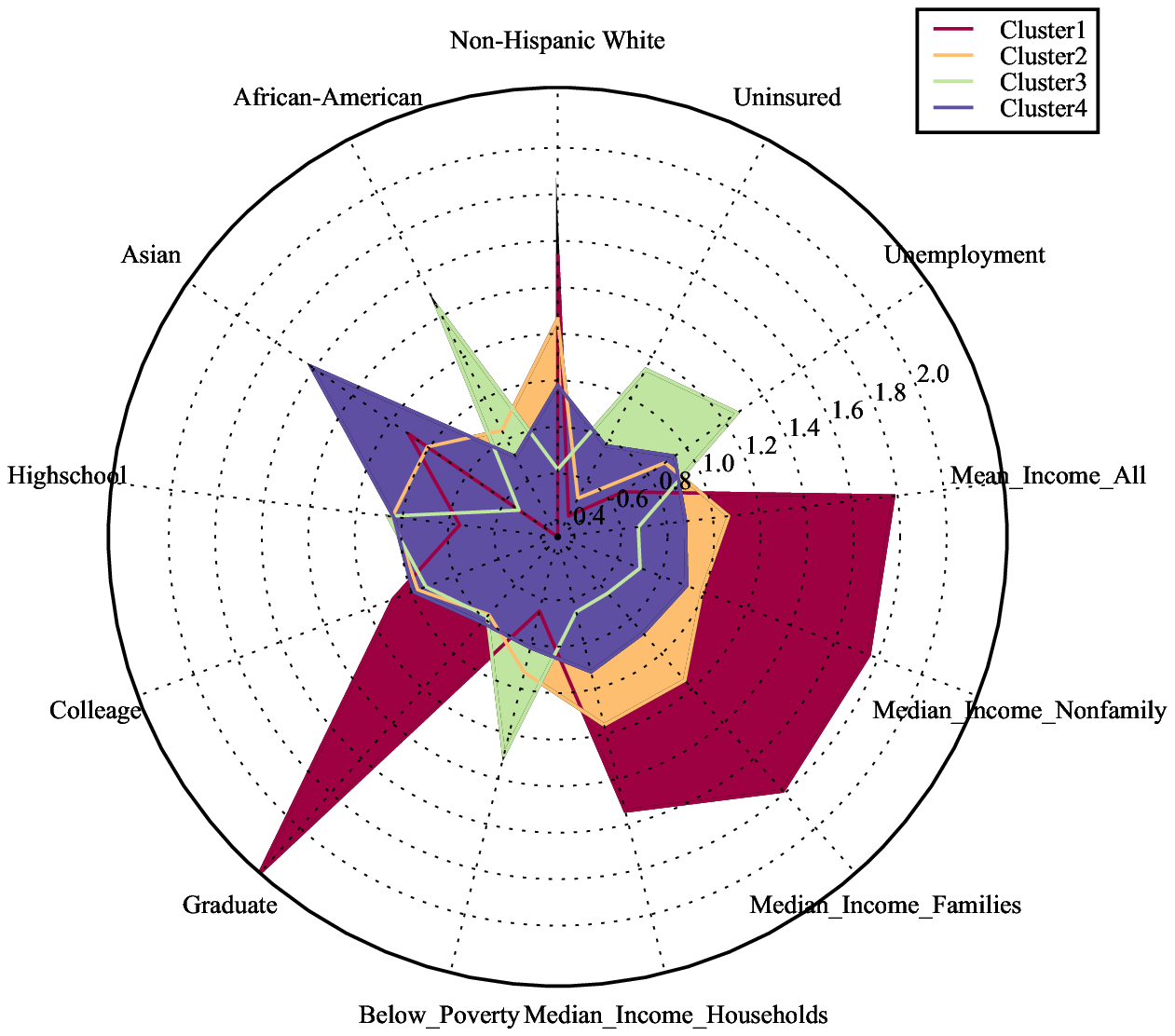}
  \label{fig:sfig2}
\end{subfigure}
\caption{Normalized ratio of socioeconomic features among clusters in New York}

\begin{subfigure}{.5\textwidth}
  \centering
  \includegraphics[height=6cm]{Fig4.pdf}
  \label{fig:sfig2}
\end{subfigure}%
\begin{subfigure}{0.5\textwidth}
  \centering
  \includegraphics[height=6cm]{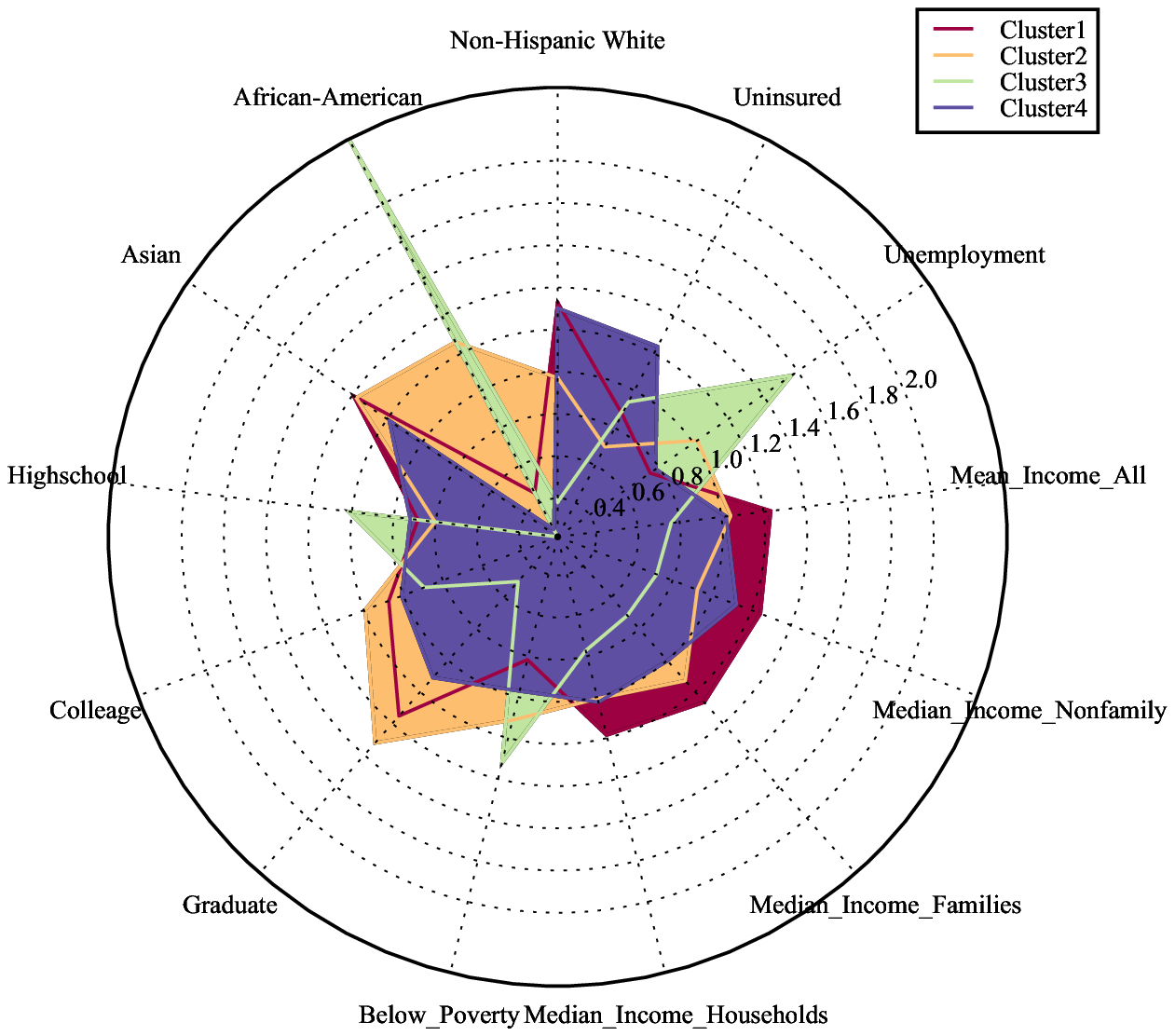}
  \label{fig:sfig2}
\end{subfigure}
\caption{Normalized ratio of socioeconomic features among clusters in Chicago}

\begin{subfigure}{.5\textwidth}
  \centering
  \includegraphics[height=6cm]{Fig3.eps}
  \label{fig:sfig2}
\end{subfigure}%
\begin{subfigure}{0.5\textwidth}
  \centering
  \includegraphics[height=6cm]{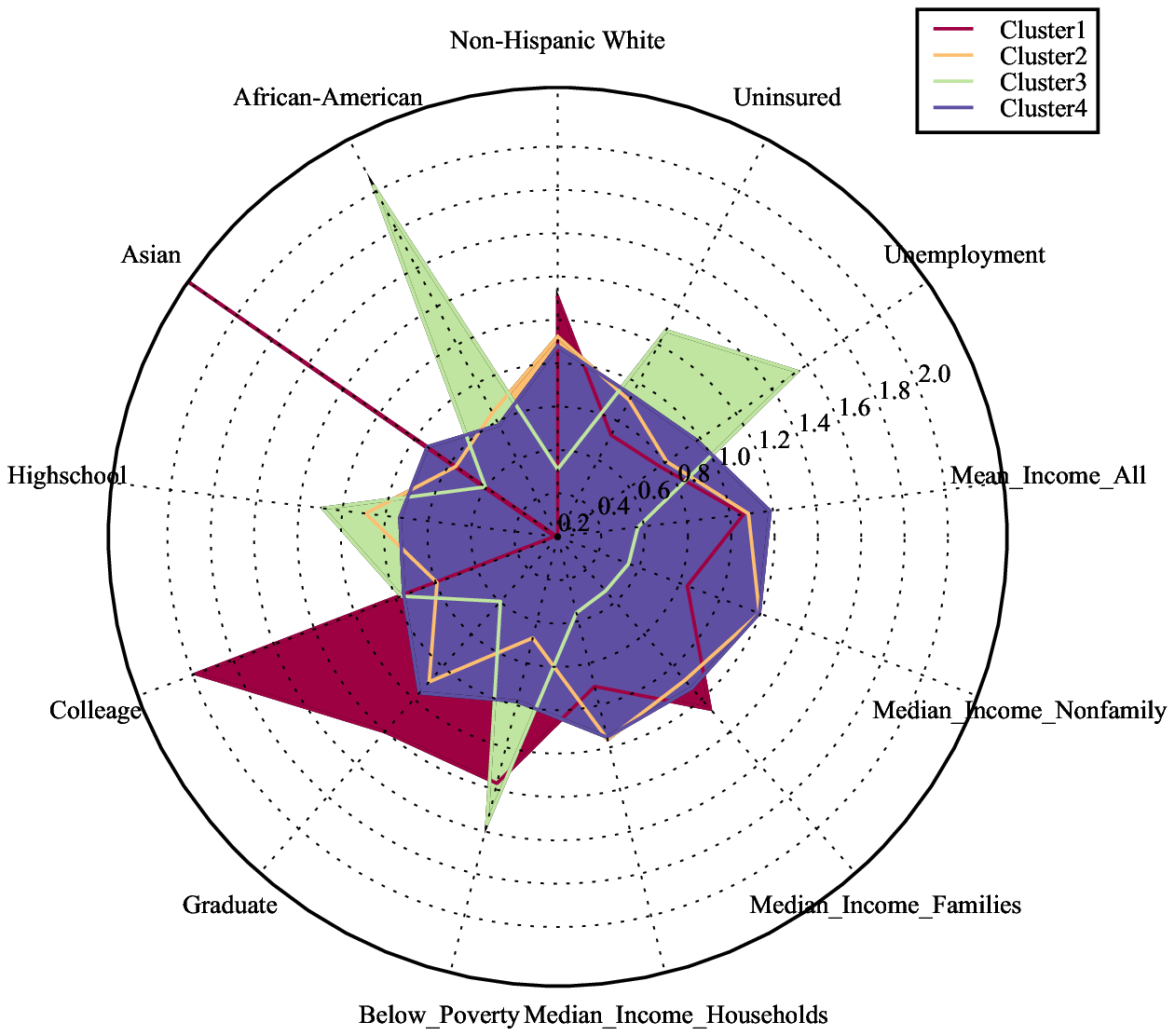}
  \label{fig:sfig2}
\end{subfigure}
\caption{Normalized ratio of socioeconomic features among clusters in Boston}
\label{fig:fig}
\end{figure}

\section{Modeling the socioeconomic features}

We find that 311 service request signatures allow the city to be divided into clusters based on distinctive patterns of socioeconomic characteristics. Following this, we explore whether 311 service requests can be used to model these socioeconomic patterns. Such a model could be useful as socioeconomic data are often unavailable or inconsistent at a given spatio-temporal scale, and therefore having a proxy derived from a model based on regularly-updated open data could have considerable potential for city operations and neighborhood planning.

We train regression models over the relative frequencies of 311 service requests of each type in each census tract in order to estimate the selected socioeconomic features described in subsection 1.2.1. The service requests frequencies $s(a,t)/s(a)$ (components of the signature vectors) constitute our feature space, including 179 different features in the case of NYC, across 2000 census tracts following the data cleaning/filtering process.


We consider six target variables including income per capita, percentage of residents with a graduate degree, percentage of unemployed residents, percentage of residents living below the poverty level, as well as demographic characteristics including percentage of Non-Hispanic White and African-American populations.

The objective of the modeling is to use partial information about the target variables defined in a certain part of the city to train the model so that it can explain the target variables over the rest of the city. 

For the purpose of a comprehensive model evaluation we use a cross-validation procedure. We try several models including Lasso\cite{hans2009}, Neural Networks with regularisation (NN)\cite{Haykin2009, Girosi1995, Jin2004}, Random Forests Regression (RF)\cite{liaw2002} and Extra Trees Regression (ETR)\cite{sklearn, Geurts2006}.

For each model, we treat the different set of hyper parameters as different models. For Neural Networks, we try 5, 10, 20, 40 hidden unites and for each hidden unit, we try penalization lambda for 0.0005, 0.005, 0.05, 0.5. As to the learning process, we use mini batch size 20 and we use the following learning rate and epochs: (0.1,100), (0.05,200), (0.01,500), (0.005,1000). For RF and ETR, we use 500 trees(since increasing trees does not help) and try maximum leaf nodes: 10, 20, 30, ... 100.
In total we have 64 sets of hyper parameters for NN and 10 for RF and 10 for ERF.

More details on the model selection process is presented in SI 3. Generally speaking, we select the final model with suitable hyper parameters with the help of cross-validation. We divide the data set into training and testing set by the ratio 7:3. As described above, we have 84 different models. For each model, we randomly divide our training set into training and validation sets and train the model on the training set and report the R-squared on the validation set. We repeat this process 20 times for each model and get the average R-squared. We pick the best model and use it for prediction on the test set. Finally, we report the out-of-sample R-squared in Table 2. Generally speaking, RF and ETR usually give us best performance based on R-squared.

The resulting out-of-sample R-squared for the models selected are summarized in Table 2. We consider this modeling result important because:

\begin{itemize}
    \item it indicates that a relationship exists between 311 request signature and the local socioeconomic features of each area;
    \item it enables possible prediction and estimation of other local socioeconomic features by using 311 requests data, particularly those features for which data are collected at low temporal frequency, such as Census data; and,
    \item it can be easily scaled by geographic aggregation for various research, operational, or planning purposes.
\end{itemize}


\begin{table}[ht]
\centering
\resizebox{\textwidth}{!}{\begin{tabular}{rrrrrrrrr}
  \hline
 City & White/European & Afro-American & Graduate Degree & Income per cap&Below Poverty&Unemployment\\ 
  \hline
NYC & 0.54 & 0.50 & 0.48 & 0.70 & 0.44 & 0.26 \\ 
Chicago& 0.76 & 0.85 & 0.45 & 0.55 &0.52 & 0.65 \\ 
Boston & 0.54& 0.68 & 0.26& 0.62 &0.63 & 0.36 \\ 

\hline
\end{tabular}}
\caption{Out of Sample R-squared} 
\end{table} 

\section{Prediction of the real estate prices}

Following our previous analysis, we attempt to understand the practical applicability of the prediction models. Although the findings above once again highlight a strong relation between 311 service request data and socioeconomic context of urban locations, this by itself has limited practical implications except for filling gaps in the data availability. In this section we show that 311 service request data could be also used to predict future socioeconomic variations, which may have more important practical implications for urban analytics. 

As an example, consider the annual average sale price of housing per square foot in different neighborhoods of NYC as the target variable for our prediction. Our housing price is reported by Zillow at the zip code level; therefore, we rescale our 311 service request frequencies to this spatial aggregation.

To match available housing price data, we only include those 311 service categories that were recorded consistently between 2012 and 2015. New York City has 145 of such categories, covering about 70 percent of total service requests. 

The target variable is updated annually and is available for each year from 2012 to 2015. The Zillow data cover 112 of the 145 zip codes in New York City where the density and frequency of 311 requests is sufficient to satisfy the filtering procedure described in the Data section. Thus, our sample for this prediction is based on data from 112 zipcodes.

We do not attempt to predict the absolute level of prices, but changes over time relative to the NYC mean. Our output therefore indicates how much more (less) expensive the housing price in a given zip code area is going to be compared to the average relative increase (decrease) in housing prices across NYC from the previous year. This way we define a new log-scale target variable $Y^t(z)$ in year $t$ as 
$$Y^t(z)=log(P^t(z)/P_{mean}^t)$$      
where $P^t(z)$ is the average price per square foot in zip code $z$ during the year $t$, while 
$P_{mean}^t$ is the average price per square foot across the entire city during the year $t$, estimated as
the mean of $P^t(z)$ for all the locations $z$ weighted by residential population 
of the locations used as a proxy for the locations' size.

We begin by modeling the output variable $Y^{2015}$. We train the model using 2012 and 2013 data (both - features and output variable) over the entire NYC and use 2014 data for tuning hyper-parameters, then apply it to 2015 using the features defined based on 2015 service requests. To reiterate, the feature space as before consists of the relative service requests frequencies $s(a,t)/s(a)$ , but now including only 145 categories of service requests, while the number of observations is 112 zip codes. 

We subsequently train four different machine learning regression models as before: Lasso\cite{hans2009}, Neural Networks with regularisation (NN)\cite{Haykin2009, Girosi1995, Jin2004}, Random Forests Regression (RF)\cite{liaw2002} and Extra Trees Regression (ETR)\cite{sklearn, Geurts2006}.

The results are reported in table 3 (we also include Boston and Chicago here just for comparison, although the number of zip codes in these cities is much smaller and thus the model becomes less significant).



\begin{table}[ht]
\centering
\resizebox{.5\textwidth}{!}{\begin{tabular}{||c c c c||}
  \hline
 Models & NYC & Chicago & Boston\\ 
  \hline
Lasso & 0.49 & 0.57 & 0.38 \\ 
NN(Regularized)& 0.70 & 0.65 & 0.68\\ 
RF& 0.78& 0.81 & 0.79 \\ 
ETR& 0.79& 0.90 & 0.83 \\ 

\hline
\end{tabular}}
\caption{Out of Sample R-squared} 
\end{table} 

As one can see from the table 3, we achieve reasonable predictive power, especially with RF and ETR approaching $R^2$ values of 0.80 for all three cities. 

However, note that modeling housing prices in 2015 is not our objective here, since a simplified model $Y^{2015}=Y^{2014}$ would achieve better results given the serial correlation in the time series and the relatively small year-to-year variation in price levels. Instead, we rather focus on the model's ability to predict the magnitude and direction of those fluctuations, forecasting price trends at the zip code level.


Let $Y_P^t(z)$ be the predicted value of $Y^t(z)$. We define
$D(z)=Y^{2015}(z)-Y^{2014}(z)$ as the actual tendency of relative real estate prices in the zip code $z$ and 
$D_P(z)=Y^{2015}_P-Y^{2014}_P$ as the predicted tendency of comparative housing price.

We classify the 112 zip codes of NYC into three groups based on the predicted tendency strength $D^{2015}_P$:\\
$G_{Positive}=\{z: D_P^i>m\cdot \sigma(D_P) \text{, where } i = 1,2,...,112\}$: group of areas with strong positive tendency;\\
$G_{Negative}=\{z: D_P^i<-m\cdot \sigma(D_P) \text{, where } i = 1,2,...,112\}$: group of areas with strong negative tendency;\\
$G_{Neutral}=\{z: -m\cdot\sigma(D_P)<D_P(z)<m\cdot \sigma(D_P) \text{, } i = 1,2,...,112\}$: group of areas with close to neutral tendency,\\

where $m$ is a certain threshold and $\sigma(D_P)$ indicates the standard deviation of $D_P(z)$. 

Additionally we classify the zip codes based on the actual tendency strength, i.e. let us introduce $G_{Positive}',G_{Negative}',G_{Neutral}'$ in the same way as above but replacing the estimated $D_P(z)$ with the real $D(z)$ in the corresponding. In this way, compared to defining strong tendency using predicted results, we define strong tendency by the real values and then test the performance of our model by the following indicators. 

For each group $G_{Positive},G_{Negative},G_{Neutral}$, we calculate its the normalized population weighted average value of actual $D(z)$ using the following formulae:
$$
\overline{D}_{Positive}= (\frac{\sum_{i\in G_{Positive}}D(z) \cdot N(z)}{\sum_{z=1}^{112}D(z)\cdot N(z)})/ \sigma(D(z)),
$$$$
\overline{D}_{Negative}= (\frac{\sum_{i\in G_{Negative}}D(z) \cdot N(z)}{\sum_{z=1}^{112}D(z)\cdot N(z)})/ \sigma(D(z)),
$$$$
\overline{D}_{Neutral}= (\frac{\sum_{i\in G_{Neutral}}D(z) \cdot N(z)}{\sum_{z=1}^{112}D(z)\cdot N(z)})/ \sigma(D(z)),
$$
where $N(z)$ is the population of the zip code $z$. Similarly for each of the groups $G_{Positive}',G_{Negative}',G_{Neutral}'$ we calculate the average prediction
$$
\overline{D}_{Positive}'= (\frac{\sum_{i\in G_{Positive}}'D_p(z) \cdot N(z)}{\sum_{z=1}^{112}D_p(z)\cdot N(z)})/ \sigma(D(z)),
$$$$
\overline{D}_{Negative}'= (\frac{\sum_{i\in G_{Negative}}'D_p(z) \cdot N(z)}{\sum_{z=1}^{112}D_p(z)\cdot N(z)})/ \sigma(D(z)),
$$$$
\overline{D}_{Neutral}'= (\frac{\sum_{i\in G_{Neutral}}'D_p(z) \cdot N(z)}{\sum_{z=1}^{112}D_p(z)\cdot N(z)})/ \sigma(D(z)),
$$
 
The values of those quantities for different values of the threshold ($m=0.15$, example of a very loose threshold classifying most of the predictions as strong, $m=0.35, 0.65, 1$) are reported in the Tables 4 and 5 and we can see consistent inequalities
$$
\overline{D}_{Positive}>0>\overline{D}_{Negative}
$$
and
$$
\overline{D}_{Positive}'>0>\overline{D}_{Negative}'
$$
holding for all the values of the threshold $m$, which means that our predicted trend directions are consistent with the real trends on average.







Moreover, we compare the signs of the predicted values of $D_P(z)$ for the strong predicted trends $G_{Positive}\cup G_{Negative}$ vs the ground-truth $D(z)$, as well as the actual values $D(z)$ for the strong actual trends $G_{Positive}'\cup G_{Negative}'$, reporting the accuracy ratio of predicting the correct trend direction for strong actual trends and the accuracy ratio for having strong predicted trends to reveal correct trend directions ($D_P(z)D(z)>0$). Those indicators are listed in Table 4 and Table 5 demonstrating the model's performance.

From Tables 4 and 5, we see that, for around 40 percent of strongest tendency observations or predictions (m=0.65), our prediction accuracy of a trend direction is higher than 80 percent compared to around 43/62(69\%) percent random guess baseline model in Table 4 and 31/51(60.7\%) baseline in Table 5. Moreover, in Table 4, we see that when the threshold m increases from 0.15 to 0.65, the accuracy ratio of prediction goes up from 70 percent to 82 percent, meaning that the stronger the actual trend, the more likely to achieve correct prediction. In Table 5, we see that while $m$ increases from 0.15 to 1, the accuracy ratio of prediction goes up from 72 percent to 90 percent, hence the stronger the predicted trend, the more accurately our prediction reflects the reality.




\begin{table}[ht]
\centering
\begin{tabular}{|*{13}{c|}} 
\hline
\multicolumn{1}{|c}{Threshold} & \multicolumn{6}{|c|}{m=0.15}& \multicolumn{6}{|c|}{m=0.35}\\ \hline

\multicolumn{1}{|c}{+/-:Strong Positive/Negative} & \multicolumn{2}{|c|}{+}& \multicolumn{2}{|c|}{-}& \multicolumn{2}{|c|}{Neutral}& \multicolumn{2}{|c|}{ +}& \multicolumn{2}{|c|}{-}& \multicolumn{2}{|c|}{Neutral}
\\ \hline
\multicolumn{1}{|c}{Number of Observations} & \multicolumn{2}{|c|}{23}& \multicolumn{2}{|c|}{75}& \multicolumn{2}{|c|}{14}& \multicolumn{2}{|c|}{ 20}& \multicolumn{2}{|c|}{62}& \multicolumn{2}{|c|}{30}
\\ \hline
\multicolumn{1}{|c}{$\overline{D}_{Positive}'/\overline{D}_{Negative}'/\overline{D}_{Neutral}'$} & \multicolumn{2}{|c|}{134.57}& \multicolumn{2}{|c|}{-84.28}& \multicolumn{2}{|c|}{-3.75}& \multicolumn{2}{|c|}{ 148.60}& \multicolumn{2}{|c|}{-95.41}& \multicolumn{2}{|c|}{-7.97}

\\ \hline
\multicolumn{1}{|c}{Accuracy for Strong P/N} & \multicolumn{6}{|c|}{ 0.7}& \multicolumn{6}{|c|}{0.72}\\ \hline

\multicolumn{1}{|c}{Threshold} & \multicolumn{6}{|c|}{m=0.65}& \multicolumn{6}{|c|}{m=1}\\ \hline
\multicolumn{1}{|c}{+/-:Strong Positive/Negative} & \multicolumn{2}{|c|}{+}& \multicolumn{2}{|c|}{-}& \multicolumn{2}{|c|}{Neutral}& \multicolumn{2}{|c|}{ +}& \multicolumn{2}{|c|}{-}& \multicolumn{2}{|c|}{Neutral}

\\ \hline
\multicolumn{1}{|c}{Number of Observations} & \multicolumn{2}{|c|}{19}& \multicolumn{2}{|c|}{43}& \multicolumn{2}{|c|}{50}& \multicolumn{2}{|c|}{14}& \multicolumn{2}{|c|}{24}& \multicolumn{2}{|c|}{74}

\\ \hline
\multicolumn{1}{|c}{$\overline{D}_{Positive}'/\overline{D}_{Negative}'/\overline{D}_{Neutral}'$} & \multicolumn{2}{|c|}{156.73}& \multicolumn{2}{|c|}{-114.82}& \multicolumn{2}{|c|}{-24.5}& \multicolumn{2}{|c|}{179.69}& \multicolumn{2}{|c|}{-137.11}& \multicolumn{2}{|c|}{-32.56}\\ \hline
\multicolumn{1}{|c}{Accuracy for Strong P/N} & \multicolumn{6}{|c|}{ 0.82}& \multicolumn{6}{|c|}{0.77}\\ \hline
\end{tabular}
\caption{Accuracy of discovering actual strong relative real estate price trends by the predictive model} 

\end{table} 

\begin{table}[ht]
\centering
\begin{tabular}{|*{13}{c|}} 
\hline
\multicolumn{1}{|c}{Threshold} & \multicolumn{6}{|c|}{m=0.15}& \multicolumn{6}{|c|}{m=0.35}\\ \hline

\multicolumn{1}{|c}{+/-:Strong Positive/Negative} & \multicolumn{2}{|c|}{+}& \multicolumn{2}{|c|}{-}& \multicolumn{2}{|c|}{Neutral}& \multicolumn{2}{|c|}{ +}& \multicolumn{2}{|c|}{-}& \multicolumn{2}{|c|}{Neutral}
\\ \hline
\multicolumn{1}{|c}{Number of Observations} & \multicolumn{2}{|c|}{43}& \multicolumn{2}{|c|}{58}& \multicolumn{2}{|c|}{11}& \multicolumn{2}{|c|}{ 32}& \multicolumn{2}{|c|}{42}& \multicolumn{2}{|c|}{38}
\\ \hline
\multicolumn{1}{|c}{$\overline{D}_{Positive}/\overline{D}_{Negative}/\overline{D}_{Neutral}$} & \multicolumn{2}{|c|}{22.61}& \multicolumn{2}{|c|}{-75.99}& \multicolumn{2}{|c|}{-4.56}& \multicolumn{2}{|c|}{ 42.23}& \multicolumn{2}{|c|}{-71.18}& \multicolumn{2}{|c|}{-40.78}

\\ \hline
\multicolumn{1}{|c}{Accuracy for Strong P/N} & \multicolumn{6}{|c|}{ 0.72}& \multicolumn{6}{|c|}{0.77}\\ \hline

\multicolumn{1}{|c}{Threshold} & \multicolumn{6}{|c|}{m=0.65}& \multicolumn{6}{|c|}{m=1}\\ \hline
\multicolumn{1}{|c}{+/-:Strong Positive/Negative} & \multicolumn{2}{|c|}{+}& \multicolumn{2}{|c|}{-}& \multicolumn{2}{|c|}{Neutral}& \multicolumn{2}{|c|}{ +}& \multicolumn{2}{|c|}{-}& \multicolumn{2}{|c|}{Neutral}

\\ \hline
\multicolumn{1}{|c}{Number of Observations} & \multicolumn{2}{|c|}{20}& \multicolumn{2}{|c|}{31}& \multicolumn{2}{|c|}{61}& \multicolumn{2}{|c|}{15}& \multicolumn{2}{|c|}{12}& \multicolumn{2}{|c|}{85}

\\ \hline
\multicolumn{1}{|c}{$\overline{D}_{Positive}/\overline{D}_{Negative}/\overline{D}_{Neutral}$} & \multicolumn{2}{|c|}{44.93}& \multicolumn{2}{|c|}{-70.55}& \multicolumn{2}{|c|}{-29.83}& \multicolumn{2}{|c|}{110.80}& \multicolumn{2}{|c|}{-76.29}& \multicolumn{2}{|c|}{-41.17}\\ \hline
\multicolumn{1}{|c}{Accuracy for Strong P/N} & \multicolumn{6}{|c|}{ 0.83}& \multicolumn{6}{|c|}{0.90}\\ \hline
\end{tabular}
\caption{Accuracy of the correspondence of the predicted strong relative real estate price trends to the actual ones} 

\end{table}

The results presented in this section demonstrate that the 311-based model can indeed predict future fluctuations of socio-economic characteristics, including real estate price trends. This serves as an initial proof of concept for multiple potential urban applications using 311 data as a proxy for local socio-economic conditions.

\section*{Conclusions}

A quantitative understanding of urban neighborhoods can be quite challenging for urban planners and policy-makers given significant gaps in the spatial and temporal resolution of data and data collection modalities. However, this subject is crucial for urban planning and decision making, as well as for the study of urban economic and neighborhood change. In this paper, we provide an approach to quantify local signatures of urban function via 311 service request data collected in various cities across the US. These datasets, which can be easily scaled by spatial (zip code, census tracts/blocks, etc.) and temporal level of aggregation, are open to the public and updated regularly. Importantly, we demonstrate consistent relationships between socioeconomic features of urban neighborhoods and their 311 service requests. 

For all three cities analyzed - New York City, Boston and Chicago - we demonstrate how clustering of census tracts by the relative frequency vectors of different types of 311 requests reveal distinctive socioeconomic patterns across the city. Moreover, those frequency vectors 
allow us to train and cross-validate regression models successfully explaining selected socioeconomic features, such as education level, income, unemployment and racial composition of urban neighborhoods. For example, the accuracy of the model explaining local average income in NYC is characterized by a R-squared value of 0.7, while Extra Trees Regression results in a 0.9 out of sample R-squared in explaining housing prices in Chicago (although this must be considered with respect to the smaller sample size). Finally, we illustrate the predictive capacity of the approach by training and validating the model to detect comparative average real estate price trends for zip codes in New York City.

In the nascent field of urban science and more traditional disciplines of economics and urban planning, there is increasing attention on how data collected by cities can be combined with novel machine learning approaches to yield insight for researchers and policy-makers. It is possible that such data can be used to better understand the dynamics of local areas in cities, and support more informed decision-making. In addition, it is conceivable that a set of efficient instrumental variables based on widely-available 311 data can be used to replace survey-based socioeconomic statistics at spatio-temporal scale where such official survey data is non-existent or inconsistent, thus broadening opportunities for urban analytics.



\section*{Acknowledgments}
The authors would like to thank Brendan Reilly and other colleagues at the Center For Urban Science And Progress at NYU for stimulating discussions which helped further shaping this research and manuscript.

\newpage

\section*{S1 Text. Unsupervised model and cluster number selection}
Let S be a set of observations (locations in our data set), and a clustering U on S is a way of partitioning S into non-overlapping subsets {$U_1,U_2,...,U_k$}. We will investigate how well the model performs with different number of clusters, i.e. different k.

Here we choose K-means clustering with four clusters as our basic model. We made this decision based on the two clustering evaluation methods: Silhouette method and Elbow method$^{[1]}$.\\\

\subsection*{S1.1 Silhouette method}
Silhouette is a commonly used method of interpretation and validation of consistency within clusters of data. It was first described by Peter J. Rousseeuw in 1986$^{[3]}$ and it measures how similar an object is to its own cluster (internal relation) compared to other clusters (external relation). The Silhouette score ranges from -1 to 1, where higher value indicates better match to its own cluster and, at the same time, poorer matched to neighboring clusters---hence, higher Silhouette score means a better model overall as it highlights the distinctions among clusters.

The Silhouette value can be calculated with any distance metric, such as the Euclidean distance we applied here. We have run the tests for all three cities. The results are plotted on Figure S1.1. We can see that:
\begin{itemize}
  \item New York City: Models with 2, 3, and 4 clusters seem closely comparable and outperform the rest;
  \item Boston: Models with 2 and 4 clusters have the highest Silhouette scores;
  \item Chicago: Silhouette score appears to be decreasing as the number of clusters rises, the optimal choice is 2.
\end{itemize}

\renewcommand{\thefigure}{S1.1}

\begin{figure*}[h]
\centering
\includegraphics[width=8cm, height=6cm]{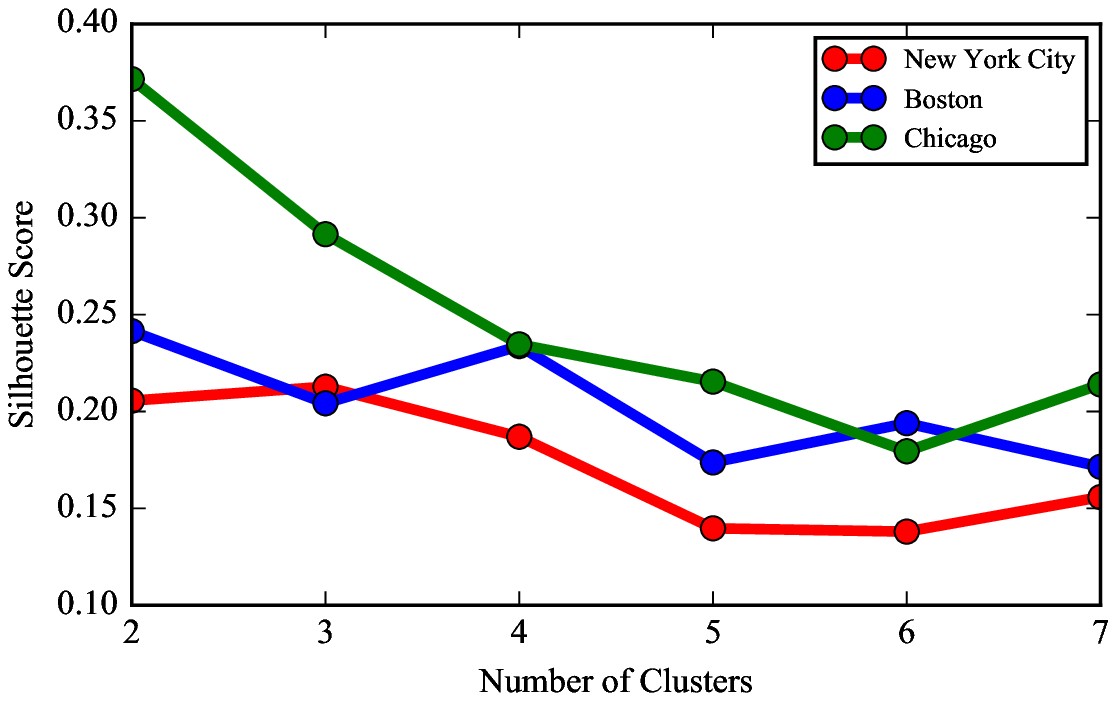}
\caption{\label{fig::time-line_classification} Silhouette method}
\end{figure*}

This observation tells us two things:
\begin{itemize}
  \item Models with 2, 3, and 4 clusters are generally better than others;
  \item 2-cluster model seems to be the best choice in terms of Silhouette's quantitative criteria.
\end{itemize}

Next we try Elbow method, another validation approach described in next subsection, before making final decisions.\\\

\subsection*{S1.2 Elbow method}

The Elbow method measures how "cost-efficient" a model is by looking at the percentage of variance explained as a function of the number of clusters. It searches for a balance between "more information" and "less complicated model". Intuitively, if we start from 1-cluster model (which is no processing at all, just leave them as a whole), adding another cluster should give more information about the data distinction, but one should stop when the marginal gain is insignificant compared to the cost. Then the number of clusters is chosen at this point$^{[5]}$.

Equivalently, we can check the average sum of squared errors. Of course, we want our error as small as possible, and the error tends to decrease toward 0 as we increase the cluster number, k (the error is 0 when k is equal to the number of data points in the dataset, because then each data point is its own cluster, and there is no error between this point and the center of its cluster--that point itself). The goal is the same: search for the point where the marginal drop is no longer attractive beyond it.

The results are summarized in Figure S1.2:

\begin{itemize}
  \item New York City: Obviously the error drops rapidly before 4 and then slows down after 4, so 4-cluster model is the best choice here;
  \item Chicago: Very similar to NYC, although the change is a bit mild and both - 3 and 4 - seem to be good choices;
  \item Boston: The trend does not provide any intuitive number of clusters to focus on.
\end{itemize}

\renewcommand{\thefigure}{S1.2}

\begin{figure*}[h]
\centering
\includegraphics[width=8cm, height=6cm]{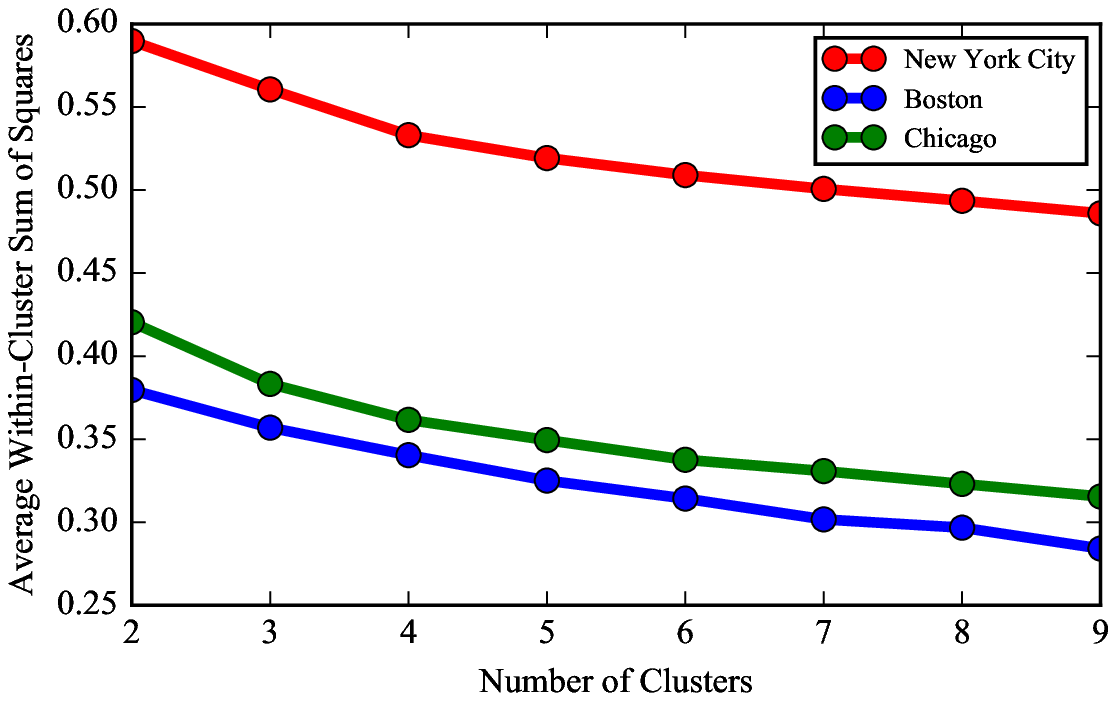}
\caption{\label{fig::time-line_classification} Elbow method}
\end{figure*}

\subsection*{S1.3 Conclusion}
To sum up, we have the following observations among three major cities in Table S1:

\setcounter{table}{0}
\renewcommand{\thetable}{S\arabic{table}}

\begin{table}
\begin{center}
\begin{tabular}{ |l|l|l| } 
 \hline
  & Silhouette & Elbow \\
 \hline
 \hline
 NYC & 3 & 4 \\
 \hline
 Chicago & 2 & 3, 4 \\
 \hline
 Boston & 2, 4 & 2 \\
 \hline
\end{tabular}
\caption{Optimal choices based on each evaluation method}
\end{center}
\end{table}

Since 4 is the only number that has appeared in all three rows, and clearly 4 clusters can reveal more details about the city structures than 2 or 3, we think that 4-cluster model may the best overall choice. Choosing 4 instead of, say, 2, in our opinion, is a reasonable trade-off between having more clusters and still decent clustering quality.


\section*{S2 Text. Classification result based on 311 service requests timeline data}

The 311 service request data is pretty rich and although types of requests considered in the paper provide an important and useful perspective for spatial clustering and modeling socio-economic quantities, there are other interesting dimensions in the data to consider. In this supplementary paragraph, we provide an alternative approach to conduct the clustering analysis based on 311 service requests data. Instead of using types of 311 service requests, we consider their timeline, building our new data set by accumulating all types of 311 services requests during each hour of the week for each zip code area. Thus this new data set includes 168 features (activity distribution per hours within an average), for all the zip code areas within New York City.

Based on the new 168 dimensional feature space, we divide the zip code areas in NYC into four clusters using K-Means clustering algorithm, highlighting different temporal patterns in 311 service request activity.  The clustering result is shown on the Figure S2.1, while the corresponding 311 service request timelines for different clusters --- on the Figure S2.2.  

\renewcommand{\thefigure}{S2.1}
\begin{figure*}[h]
\centering
\includegraphics[width=12cm, height=12cm]{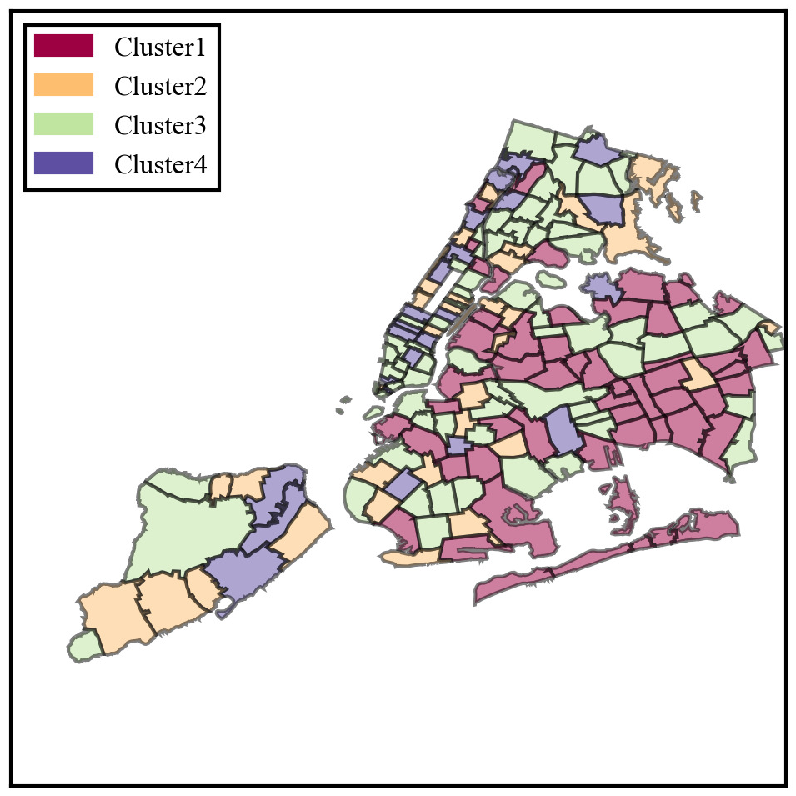}
\caption{\label{fig::time-line_classification} Classification of urban locations based on the timeline data of 311 request services}
\end{figure*}

\renewcommand{\thefigure}{S2.2}
\begin{figure*}[h]
\centering
\includegraphics[width=1.\textwidth]{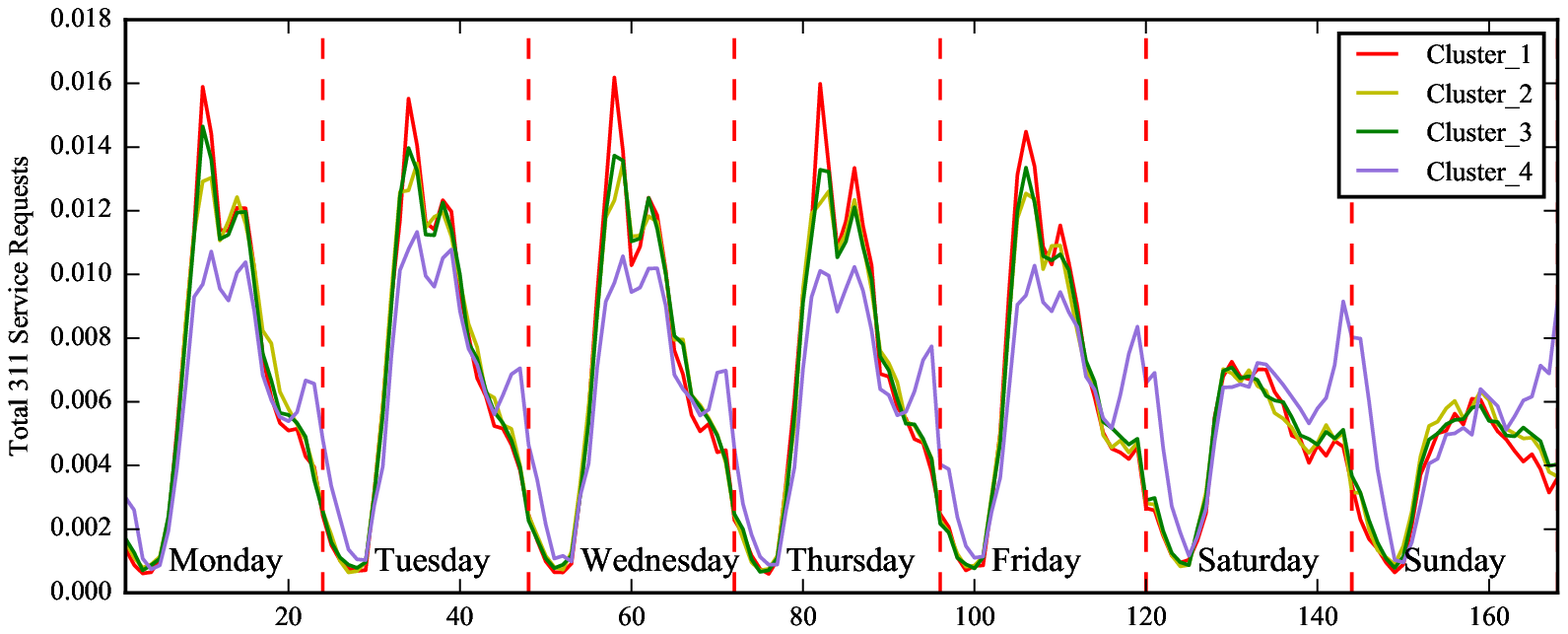}
\caption{\label{fig::time-line_classification} Hourly distribution of 311 service requests among clusters.}
\end{figure*}

Figure S2.2 shows that the timelines of 311 service requests among Clusters 1,2,3 are rather similar. But Cluster 4 can be distinguished since the service requests have a considerable spike after 8 PM each day and especially over the weekends, indicating evening-time activity in those areas, which largely include locations across Manhattan, which makes common sense. Further understanding this pattern might require additional analysis of the service requests' types and other contextual information. 
However the overall difference between socio-economic factors among different clusters is much less significant than the result based on 311 service request type data shown in Figure 4. 
Therefore, for the purpose of the socio-economic analysis of this study we decided to stick to the service request types other the timeline.

\section*{S3 Test. Model selection, Cross-validation and Out of Sample R-squared}
The results of the model selection are shown in Table 2. The following procedures are applied for each city. 

Assuming we have data set $X$ and labels $y$, where $X$ is $n\times m$ matrix and $y$ is $n\times 1$ vector. All the entries are real values. 

Firstly, we collect all the possible models for training. The types of the models we consider include: Lasso, Neural Networks with regularization, Random Forest Regression, and Extra Trees Regression. We treat models with different hyper-parameters set as different models even they are belong to the same model type. 

Secondly, we randomly split the whole data set into training data and test data. We train each model's parameters on training set, and get the out of sample R-squared from the prediction result on testing set. We repeat this process ten times, and record the average R-squared for each model.

Finally, we report the largest out of sample R-squared among all the records(models) from second part and write it in Table 2.


\section*{S4 Test. Choice of scale among zip code, census tract and census block.}

Consider the spatial granularity for the spatial aggregation of the New York City's 311 service requests, choosing among the three options: zip code areas, census tract areas, and census block areas of New York City.

The primary goal for this scale selection is to find the right balance between the number of spatial units which will serve as observations for our model and the sparsity of the data. In Figure S4, we use x-axis for the number of total requests in each area (zip code level, census block level, etc), and for each given $x$ show the number $y$ of areas with request activity higher than $x$. We hope to find an appropriate scale such that it provides both adequate number of areas to analyze and abundant request activities to analyze per each area.
\begin{itemize}
    \item Let's start with Zip Code scale--only 178 total observations at hand, it's too few to apply various machine learning algorithms for our research, despite most zip code areas have more than 500 activities in total.
    \item Census Block scale, on the other hand, offers more than 6000 areas in total. But for most of these areas (93\%), total activities are less than 500.
    \item In comparison, Census Tract data set has 1367 observations with more than 500 requests, which seems to be a good balance addressing the issue of data sparsity as well as providing enough areas to analyze.
\end{itemize}
Hence we have selected Census Tract as the basic spatial scale for our research.

\renewcommand{\thefigure}{S4}

\begin{figure*}[h]
\centering
\includegraphics[width=12cm, height=12cm]{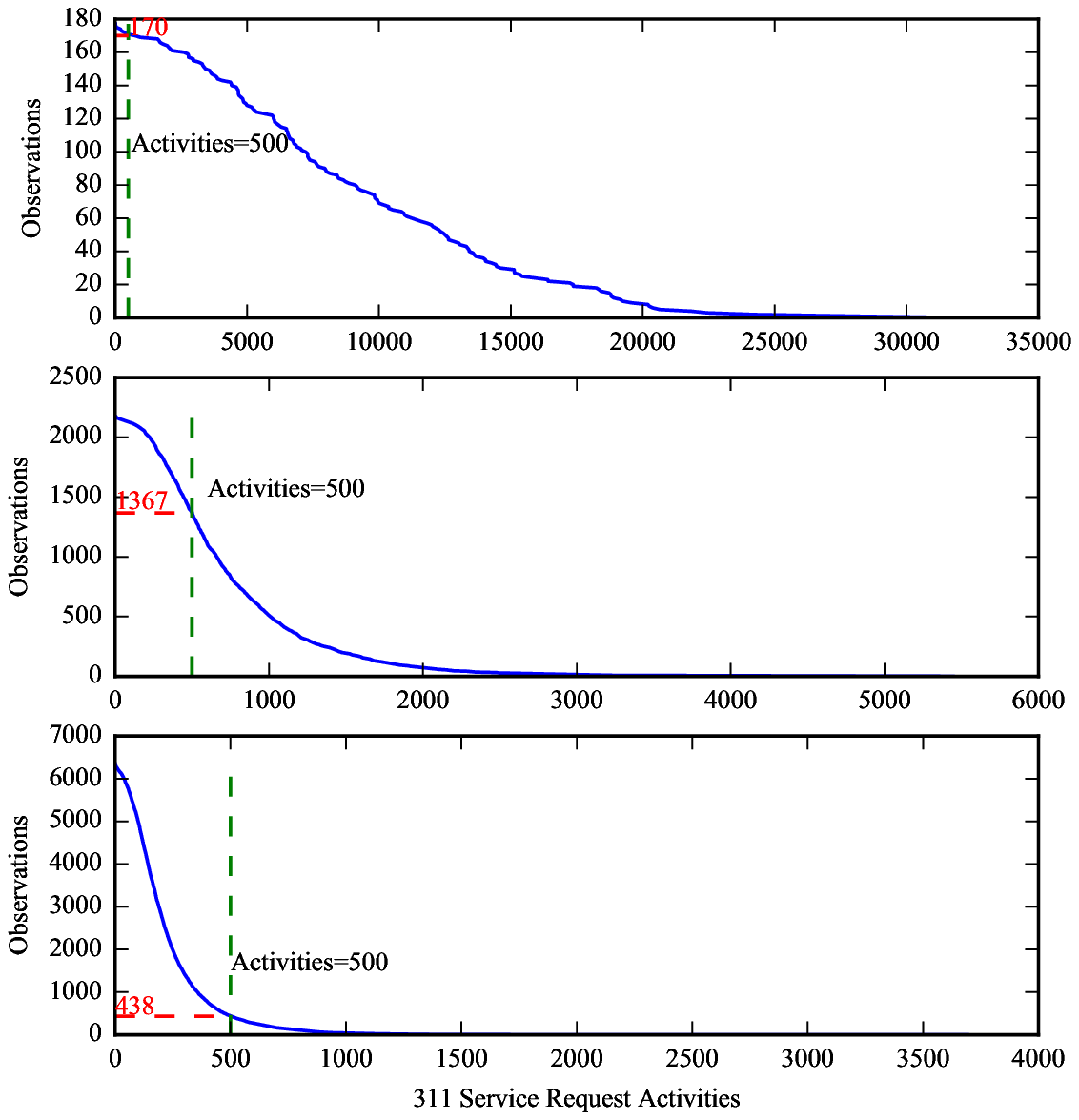}
\caption{\label{fig::time-line_classification} Number of areas vs Request Activity per Area}
\end{figure*}

\end{document}